\documentclass[%
 reprint,
 amsmath,amssymb,
 aps,
 onecolumn,
 prf,
]{revtex4-1}
\usepackage{longtable}
\usepackage[table,x11names]{xcolor}
\usepackage{graphicx}
\usepackage{dcolumn}
\usepackage{bm}
\usepackage[shortlabels]{enumitem}
\usepackage{tikz}
\usetikzlibrary{external}
\usepackage{pgfplots}
\pgfplotsset{compat=1.14}
\usetikzlibrary{matrix}
\usetikzlibrary{patterns}
\usepgfplotslibrary{fillbetween}
\usepackage[percent]{overpic}

\begin{document}

\preprint{APS/XXX-PRF}

\title{Two-dimensional dynamics of elasto-inertial turbulence \\ and its role in polymer drag reduction}
\author{S.~Sid}
\author{V.E.~Terrapon}
\email{vincent.terrapon@ulg.ac.be}
\affiliation{Department of Aerospace and Mechanical Engineering,  University of Li\`{e}ge, Belgium}
\author{Y.~Dubief}
\affiliation{School of Engineering, University of Vermont, VT 05405, USA}
\date{\today}

\begin{abstract}
The goal of the present study is threefold: (i) to demonstrate the two-dimensional nature of the elasto-inertial instability in elasto-inertial turbulence (EIT), (ii) to identify the role of the bi-dimensional instability in three-dimensional EIT flows and (iii) to establish the role of the small elastic scales in the mechanism of self-sustained EIT.
Direct numerical simulations of FENE-P fluid flows are performed in both two- and three-dimensional straight periodic channels. The Reynolds number is set to $\mathrm{Re}_\tau$ = 85 which is sub-critical for two-dimensional flows but beyond transition for three-dimensional ones. The polymer properties selected correspond to those of typical dilute polymer solutions and two moderate Weissenberg numbers, $\mathrm{Wi}_\tau$ = 40, 100, are considered. 
The simulation results show that sustained turbulence can be observed in two-dimensional sub-critical flows, confirming the existence of a bi-dimensional elasto-inertial instability. The same type of instability is also observed in three-dimensional simulations where both Newtonian and elasto-inertial turbulent structures co-exist. Depending on the $\mathrm{Wi}$ number, one type of structure can dominate and drive the flow. For large $\mathrm{Wi}$ values, the elasto-inertial instability tends to prevail over the Newtonian turbulence. This statement is supported by (i) the absence of the typical Newtonian near-wall vortices and (ii) strong similarities between two- and three-dimensional flows when considering larger $\mathrm{Wi}$ numbers.
The role of the small elastic scales is investigated by introducing global artificial diffusion (GAD) in the hyperbolic transport equation for polymers. The aim is to measure how the flow reacts when the smallest elastic scales are progressively filtered out. The study results show that the introduction of large polymer diffusion in the system strongly damps a significant part of the elastic scales that are necessary to feed turbulence, eventually leading to the flow laminarization. A sufficiently high Schmidt number (weakly diffusive polymers) is necessary to allow self-sustained turbulence to settle. Although EIT can withstand a low amount of diffusion and remains in a non-laminar chaotic state, adding a finite amount of GAD in the system can have an impact on the dynamics and lead to important quantitative changes, even for Schmidt numbers as large as $10^2$.
\end{abstract}
\maketitle

\section{\label{sec:Intro}Introduction}
The addition of small quantities of polymer macromolecules in wall-bounded flows is known to be one of the most efficient drag reduction strategy in turbulent flows \citep{Toms1948}. Although great advances have been made in the comprehension of the mechanism of polymer drag reduction, some key properties are still poorly understood. One is the longstanding question of the nature, Newtonian or Non-Newtonian, of the maximum drag reduction (MDR) state, observed by \citet{Virk1975}. A second, possibly related to the first, is the role of a recently discovered turbulent state, elasto-inertial turbulence \citep{Samanta2013,Dubief2013,Terrapon2014}, in the dynamics of drag reduced flows. This letter first focuses on the latter through an investigation of the dimensionality and role of small scale dynamics of EIT in channel flows at a marginally supercritical Reynolds number for three-dimensional flows. The findings then lend themselves to a discussion of the possible connection between EIT and MDR.

Concerted numerical and experimental research efforts in the last two decades have led to a solid understanding of how polymers reduce skin-friction drag in wall-bounded flows. The dampening of quasi-streamwise vortices, the engine of wall turbulence \citep{Jimenez1999}, reported in many studies  \citep{Dimitropoulos2001,PTASINSKI2003,Min2003a,Min2003}, arises from the stretching of polymers in the vertical motions caused by quasi-streamwise vortices \citep{Terrapon2004}; the stretched polymers work against these motions \citep{Dubief2004,Kim2008}. These advances owe a great deal to direct numerical simulation thanks to a constitutive viscoelatic model, FENE-P (Finitely Extensible Nonlinear Elastic- Peterlin), which has been shown to capture the main dynamics of drag reduced flows with polymer additives.

Recently, the same model was at the center of the discovery of Elasto-Inertial Turbulence (EIT). EIT \citep{Samanta2013,Dubief2013,Terrapon2014} is a turbulent state driven by cyclic flow-polymer interactions, and can be found over a wide range of Reynolds numbers, from sub-critical to super-critical. This polymer-driven turbulence is akin to elastic turbulence \citep{Groisman2000,Groisman2001,Groisman2004}, a chaotic state found in inertial-less polymer flows undergoing curved mean streamlines. Beyond the sheer curiosity that the existence of EIT in parallel shear flows over a wide range of Reynolds number may elicit, the dynamical significance of EIT in polymer flows is yet to be established. 

The first property is the dimensionality of EIT. Low-Reynolds elastic turbulence has been previously observed in two-dimensional Kolmogorov flows \citep{Berti2010}, demonstrating the ability of elastic fluctuations to generate turbulence in 2D planar shear flows. The coherent structures observed in 3D EIT channel flows \citep{Samanta2013,Dubief2013,Terrapon2014} also suggest that EIT may be 2D. Flow coherent structures depicted by the second invariant $Q$  of the velocity gradient tensor are best described as cylindrical with a definite spanwise alignment, whereas the polymer field shows that stretched polymers are confined in sheets of large streamwise and spanwise scales and very small thickness in the wall-normal direction. EIT may be fundamentally two-dimensional as the thin sheets seems to result from the stretching and the advection of polymers by the streamwise and wall-normal velocity components. If the two-dimensionality of EIT is confirmed, it would imply that the underlying instability exists in three dimensional flows and does not require any of the three-dimensional instabilities known in Newtonian flows to exist. In this letter, we therefore conduct simulations in two and three dimensions for the same flow and polymer parameters in order to investigate the survival of EIT against Newtonian 3D instabilities.

The second focus of our study is the role of small scales in the dynamics of EIT. We have long suspected that small scales matter in turbulent polymer flows \citep{Dubief2005}, however it so far remains unclear whether the small scales of the polymer field have any significance onto the dynamics of the flow and therefore onto polymer drag reduction. The simulation of viscoelastic flows using the Eulerian framework is achieved by coupling the transport equation for the polymer conformation tensor to the Navier-Stokes equations. One important property of this transport equation is that there is no diffusive term and the terms other than the material derivative are not diffusive in nature \citep{Dubief2005}. Thus, from a mathematical point of view, the equation is hyperbolic. The absence of diffusive term can be justified by the very low diffusion of polymers in solvents which has been reported to be of the order of $10^8$ cm$^2$/s \citep{Layec1983}. For dilute polymer solutions, it translates into Schmidt numbers ($\mathrm{Sc}$), defined as the ratio between the solvent viscosity and the polymer molecular diffusivity, of the order of $10^6$. The justifiable neglect represents a critical modeling challenge for direct numerical simulation of polymer flows by making the transport equation a stiff hyperbolic problem. There are two different approaches to solve this numerical challenge. The first is to add an artificial diffusive term to the transport equation to stabilize the simulation. However, a very low Schmidt number, $\mathrm{Sc} \approx 0.5$, has typically been used in practice \citep{Sureshkumar1997,PTASINSKI2003,DeAngelis2003,Stone2004,Xi2010,Li2006}. The second relies on techniques used in compressible or multiphase flows to handle sharp discontinuities (shocks, gas-liquid interface) that attempt to add diffusion only in regions where gradients become too large to be solved on the local meshes \citep{Min2001,Dubief2005}. This approach, akin to implicit large eddy simulation \citep{Dubief2005}, allows to model with higher fidelity the dynamics of $\mathrm{Sc}=\infty$ in the range of resolved scales. To date, the first approach has not demonstrated an ability to simulate EIT, whereas EIT was discovered by the second approach. The comparison between the two approaches set the stage for a unique experiment to study the effects of small scales on EIT and polymer drag reduction by varying the Schmidt number.

According to the theory of passive scalar transport in turbulent flows \citep{Batchelor1959,Batchelor1959b}, the smallest scalar scale is a function of the smallest scale of the flow and the Schmidt number. For a passive scalar in a turbulent flow of smallest scale $\eta_{\text{K}}$ (Kolmogorov scale), the smallest scale of the scalar is the Obukhov-Corrsin scale $\eta_{\text{OC}}=\eta_{\text{K}} \mathrm{Sc}^{-3/4}$ when $\mathrm{Sc}<1$ and the Batchelor scale, $\eta_{\text{B}}=\eta_{\text{K}} \mathrm{Sc}^{-1/2}$ when $\mathrm{Sc} \gg 1$. Since polymers are not passive but active scalar, one may not expect that Batchelor's theories directly and completely apply to our system. However, as discussed earlier \citep{Dubief2005}, there is reasonable ground to anticipate that polymers should share some key dynamical features with high Schmidt number scalar transport. Consequently, we study the range of Schmidt numbers over which EIT is observed and the effects of finite Schmidt numbers on the dynamics of the flow and polymer fields.

Finally, it is worth noting that particular care has been taken to ensure the quality of the results reported in this paper. In order to verify the adequacy of the simulation methodology, two codes using different numerical schemes have been developed. The dynamical properties of the flows obtained using the two approaches have been compared through statistical distributions of mean solvent velocity and polymer elongation. Moreover, a thorough mesh convergence study has been carried out to determine the spatial resolution necessary to properly capture all the relevant scales of the flow. The numerical analysis demonstrated that for a sufficiently fine grid, the solution is quasi-independent of the discretization for both finite and infinite Schmidt numbers. The discussion related to the simulation results accuracy is available in the supplementary material attached to this paper.

\section{\label{sec:Setup}Simulation methodology}
Direct numerical simulations of EIT are performed in two- and three-dimensional straight periodic channels for different flow conditions. The incompressible form of the Navier-Stokes equations coupled with a transport equation for the conformation tensor $\mathbf{C}$ are solved over the computational domain. The system can be written under its non-dimensional form as,
\begin{eqnarray}
	\nabla \cdot \mathbf{u} &=& 0 \nonumber \ , \\
	\partial_t \mathbf{u} + \mathbf{u} \cdot \nabla \mathbf{u}&=& - \nabla p + \frac{\beta}{\mathrm{Re}_\tau} \nabla^2 \mathbf{u} + \frac{1 - \beta}{\mathrm{Re}_\tau} \nabla \cdot \boldsymbol\tau^p + f_x \mathbf{e}_x \ , \label{eq:NS}\\
	\partial_t \mathbf{C} + \mathbf{u} \cdot \nabla \mathbf{C}&=& \nabla \mathbf{u} \cdot \mathbf{C} + \mathbf{C} \cdot \left( \nabla \mathbf{u} \right)^\mathrm{T} - \boldsymbol\tau^p + \frac{1}{\mathrm{Re}_\tau \mathrm{Sc}} \nabla^2 \mathbf{C} \ , \nonumber
\end{eqnarray}
where $\boldsymbol \tau^p$ is the polymer stress defined using the FENE-P model,
\begin{equation}
	\boldsymbol\tau^p = \frac{1}{\mathrm{Wi}} \left[ \frac{\mathbf{C}}{1 - \mathrm{tr}\left( \mathbf{C}\right)/L^2} - \mathbf{I} \right] \ ,
	\label{eq:FENEP}
\end{equation}
in which polymer molecules are modeled as dumbbells composed of two beads and a non-linear spring. The conformation tensor $\mathbf{C} = \left\langle \mathbf{q} \otimes \mathbf{q}\right\rangle$ is defined as the local averaging of the tensor product between dimensionless end-to-end vectors $\mathbf{q}$ that characterize the orientation and the length of each dumbbell. The governing equations are controlled by five non-dimensional parameters; the Reynolds number $\mathrm{Re}_\tau = u_\tau \delta / \nu$ based on the friction velocity $u_\tau$, the half channel width $\delta$ and the zero-shear-rate kinematic viscosity of the solution $\nu$; the ratio between the solvent and solution viscosity $\beta = \nu^s/\nu$; the Schmidt number $\mathrm{Sc} = \nu^\text{s}/\alpha$ defined as the ratio between the solvent kinematic viscosity $\nu^\text{s}$ and polymer diffusivity $\alpha$; the Weissenberg number $\mathrm{Wi} = \lambda u_\tau / \delta$ which compares the polymer relaxation time $\lambda$ to the wall shear rate of the corresponding Newtonian flow $u_\tau / \delta$; and the maximum dimensionless polymer extension $L$. Note that the viscous Weissenberg number defined as $\mathrm{Wi}_\tau = \mathrm{Wi} \ \mathrm{Re}_\tau = \lambda u_\tau^2 / \nu$ is the one used in this paper to define the flow conditions. The mean flow is driven by the body force $f_x \mathbf{e}_x$ which is a constant pressure gradient along the streamwise direction.

The non-dimensional parameters selected for the study are the following: $\mathrm{Re}_\tau$ = 84.95, $\beta$ = 0.97, $L$ = 70.9, $\mathrm{Wi}_\tau$ = $\left[ 40, 100 \right]$. A friction Reynolds number of $\mathrm{Re}_\tau$ = 84.95 is on the lower end of the low-Reynolds number turbulent regime for channel flows and just beyond the critical value of $\mathrm{Re}_{\tau,c} \approx$ 60 for which Newtonian instabilities appear in three-dimensional flows. However, for two-dimensional channels, the critical bulk Reynolds number is approximatively $\mathrm{Re}_{b,c} \approx$ 5772 \citep{Schmid2012} which is larger than a value of $\mathrm{Re}_{b}$ = 4817 corresponding to $\mathrm{Re}_\tau$ = 85. Lastly the choice of Reynolds number, domain dimensions, resolution, and polymer parameters allows for a direct comparison with DNS of viscoelastic flows using $\mathrm{Sc}\leq 0.5$ \citep{Xi2010,Xi2010a,Xi2012}.

The $\beta$ parameter is selected in the typical range for dilute polymer solutions and the maximum dimensionless polymer extension $L$ is chosen sufficiently large to allow the solvent shear to stretch the polymer chains. The Schmidt number is set to infinite (no diffusive term) for the simulations used to investigate the elasto-inertial instability in order to prevent excessive polymer diffusion to affect the flow solution and allow all the relevant scales to freely develop. Respecting the hyperbolic nature of the polymer transport model is key to accurately simulate viscoelastic turbulence as it will be shown in the third part of the results section.

Viscoelastic flows are stable according the the linear stability theory~\citep{Zhang2013}. Thus, an initial perturbation is necessary to trigger turbulence. The present methodology uses an alternated blowing-suction transervse excitation \cite{Dubief2013} added to the laminar viscoelastic Poiseuille flow during a short period of time to perturb the stable solution. After several flow-through times, the external excitation is turned off and the flow naturally develops towards its turbulent statistically steady state.

The channel length is chosen as long as possible to minimize possible numerical effects due to short-range interactions between highly stretched polymers through the periodic boundaries. A length of $L_x^+ = L_x u_\tau / \nu = 720$ is selected for all the simulations and three-dimensional simulations use a span of $L_z^+ = L_x u_\tau / \nu = 255$. It should be noted that EIT exists in domains with half the streamwise dimension of the present computational domain.

The numerical resolution of the system of equations (\ref{eq:NS}) is performed using second order finite difference schemes with a staggered variable arrangement \citep{Desjardins2008, Morinishi1998} which ensures the discrete conservation of mass, momentum and kinetic energy. The continuity equation is enforced through a fractional step method \citep[][]{Kim1985a} where a Poisson equation for pressure is solved in the Fourier-physical space. The system is advanced in time using the Crank-Nicholson semi-implicit scheme. As mentioned before, the choice of the discretization and the adequacy of the numerical methods have been verified using code comparison and a mesh sensitivity analysis that is reported in the supplementary material attached to this paper.

\section{\label{sec:Results}Results}
\begin{figure}[!ht]
\centering
\includegraphics[scale=0.95,trim=30 630 50 52,clip,tics=10]{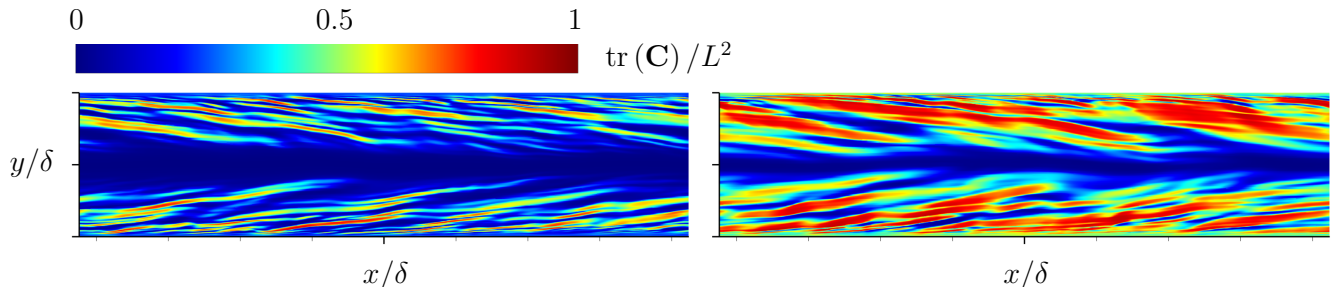}
\caption{Instantaneous contours of polymer elongation $\mathrm{tr} \left( \mathbf{C} \right) / L^2$ in two-dimensional simulations. Left: $\mathrm{Wi}_\tau = 40$, right: $\mathrm{Wi}_\tau = 100$.}
\label{fig:2D_contours}
\end{figure}

Figure~\ref{fig:2D_contours} shows the instantaneous contours of polymer elongation observed in two-dimensional elasto-inertial turbulence at the statistically steady state.
After a sufficiently long simulation time, the flow fully develops into a self-sustained chaotic motion. For the two Weissenberg numbers considered, the elastic and turbulent kinetic energies fluctuate around their mean values and do not show any sign of decay for more than 100 flow-through times (not shown here). The existence of a sustained turbulent motion in two-dimensional flows demonstrates the ability of FENE-P fluid flows to generate turbulence for sub-critical (2D) Reynolds number in wall-bounded planar configurations. 

The physical mechanism behind the elasto-inertial instability observed is very similar to the one depicted in~\cite{Dubief2013} for a similar range of $\mathrm{Wi}$ number in three-dimensional flows. Thin sheets of large polymer extension arise from the stretching and the advection of polymer macromolecules by the solvent. The accumulation of anisotropic elastic stress in the sheets, in turn, induces velocity fluctuations organized as a train of alternating circular regions of rotational and extensional topology. The flow inertia of this chaotic motion feeds the polymer stretching mechanism and regenerate polymer sheets, closing the loop of self-sustained EIT. As the flow is confined in a two-dimensional channel and is thus sub-critical, the typical Newtonian turbulent structures are not able to flourish and the dynamics is exclusively driven by the elasto-inertial instability. In fact, it has been shown that, for Newtonian fluids, reducing the channel span below a minimum size necessary to allow fundamental structures to develop hinders the cycle of near-wall turbulence and leads to the flow laminarization \citep{Jimenez1991,Jimenez1999}.

Comparing the two flow conditions indicates that, for the same level of solvent inertia (Reynolds number), increasing the Weissenberg number allows polymer chains to reach larger elongations over a wider fluid layer. In addition, the kinetic energy of momentum fluctuations also increases as they take source in the elastic stress gradients that are directly related to the local variations of polymer elongation. However, the dynamical behavior is very similar for the two $\mathrm{Wi}$ numbers considered.

\begin{figure}[!ht]
\centering
\includegraphics[scale=0.86,trim=30 500 50 52,clip,tics=10]{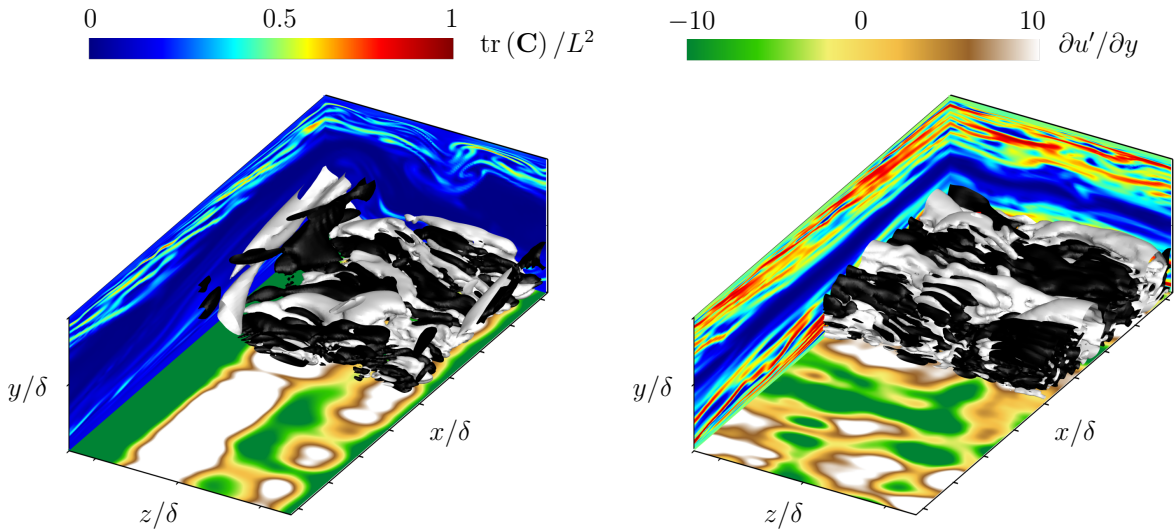}
\caption{Visualization of flow structures in three-dimensional flows. Contours of polymer elongation $\mathrm{tr} \left( \mathbf{C} \right) / L^2$ in vertical planes, contours of streamwise shear fluctuations in a near-wall horizontal plane and turbulent structures identified by large positive (white) and negative (black) $Q$-criterion values (only shown in the bottom half of the channel). Left: $\mathrm{Wi}_\tau = 40$, right: $\mathrm{Wi}_\tau = 100$.}
\label{fig:3D_contours}
\end{figure}

The flow structures observed in three-dimensional channels for the same flow conditions are reported in Figure~\ref{fig:3D_contours}.
As the Reynolds number is supercritical for three-dimensional flows, the viscoelastic turbulence observed results from the interactions between Newtonian and elasto-inertial turbulent structures. For the lower Weissenberg number, the two types of structures are clearly visible in the flow. The contours of polymer elongation in vertical planes show polymer sheet structures that are similar to those observed in two-dimensional flows. However, the sheets are not homogeneous along the channel span and are intermittently subject to the shear and advection of near-wall Newtonian vortices (see the vortical structure at the channel centerline near the top wall). These vortices are associated to long velocity streaks identified by large streamwise shear fluctuations. For the channel span selected, a single streak is observed in the flow. Despite the polymer-solvent interactions, the large streak colored in white remains coherent along the streamwise direction demonstrating the relative importance of Newtonian turbulence at this Weissenberg number. The turbulent structures identified by positive values of the $Q$-criterion \citep{Dubief2000} (rotational) also show that near-wall hairpin vortices are able to develop in the flow but are less frequent than in Newtonian turbulence. The spanwise aligned structures occupying a large part of the channel span correspond to the train of alternating circular regions of rotational and extensional topology specific to the elasto-inetial instability.

\begin{figure}[!ht]
\centering
\includegraphics[scale=0.85,trim=30 540 20 52,clip,tics=10]{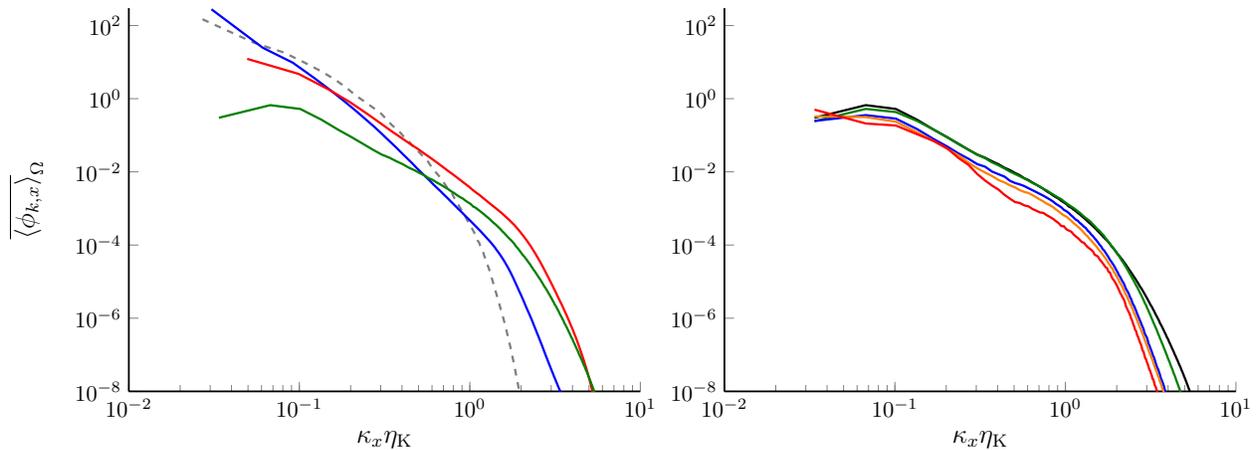}
\caption{Streamwise spectra of turbulent kinetic energy $k$ averaged over the computational domain and over time. Left: Two- and three-dimensional flows at $\mathrm{Sc} = \infty$ for different Weissenberg numbers: \textcolor{gray}{3D, Newtonian (dashed)}; \textcolor{blue}{3D, $\mathrm{Wi}_\tau$ = 40}; \textcolor{red}{3D, $\mathrm{Wi}_\tau$ = 100} and \textcolor{black!50!green}{2D, $\mathrm{Wi}_\tau$ = 40}. Right: Two-dimensional flow at $\mathrm{Wi}_\tau$ = 40 for different Schmidt numbers: \textcolor{red}{$\mathrm{Sc}$ = 9}, \textcolor{orange}{$\mathrm{Sc}$ = 16}, \textcolor{blue}{$\mathrm{Sc}$ = 25}, \textcolor{black!50!green}{$\mathrm{Sc}$ = 100} and $\mathrm{Sc} = \infty$.}
\label{fig:Spectra}
\end{figure}

At larger Weissenberg number, the picture is relatively different. The Newtonian structures are no longer observed in the flow and the dynamics appears to be dominated by the elasto-inertial instability. Near-wall vortices and velocity streaks are inhibited by a stronger elasto-inertial turbulence. The streaks have lost their coherency and the only structures that can be observed are those specific to EIT. These structures are well-organized and occupy the full channel span. The polymer sheets are homogeneous along the spanwise direction. Although additional fluctuations inherent to the extension to three-dimensional flows exist, large $\mathrm{Wi}$ EIT shares many dynamical features with two-dimensional elasto-inertial flows as this is the dominating instability. The similarities between two- and three-dimensional flows at larger $\mathrm{Wi}$ can also be measured quantitatively by comparing the power spectral density of the turbulence in the different flows. The averaged spectra analyzed are defined as,
\begin{equation}
	\overline{\left\langle \phi_{k,x} \right\rangle}_\Omega = \frac{1}{T_s} \frac{1}{\Omega} \int_{t_0}^{t_0+T_s} \int_\Omega \phi_{k,x} \left( \kappa_x, y, z, t \right) \ \mathrm{d}\Omega \ \mathrm{d}t
\quad \text{with} \quad
	\phi_{k,x} \left( \kappa_x, y, z, t \right) = \frac{1}{2 u_\tau^2} \left[ \hat{\mathbf{u}} \left( \kappa_x, y, z, t \right) \cdot \hat{\mathbf{u}}^* \left( \kappa_x, y, z, t  \right) \right] \,,
\end{equation}
where $T_s$ the time averaging window, and $\Omega$ the channel volume. The hat operator stands for the Fourier transform along the streamwise $x$-direction, $\hat{\bullet} = \mathcal{F}_x \left( \bullet \right)$, and the star superscript denotes the complex conjugate. These spectra are reported in the left sub-figure of Figure~\ref{fig:Spectra}. 
The wavenumber $\kappa_x$ is normalized by the wall Kolmogorov scale $\eta_\text{K} = \left(\nu^3 / \epsilon_\text{wall} \right)^{1/4}$ computed \textit{a posteriori} from simulation results. One can see that for the larger Weissenberg number, the spectrum of the three-dimensional flow has a shape that is closer to the one obtained for the two-dimensional flow. The small-scale content increases with $\mathrm{Wi}$ and the slope of the inertial range tends to converge towards the one obtained for the two-dimensional flow. This additional comparison corroborates previous qualitative results and demonstrate the importance of the elasto-inertial instability in moderate/large Weissenberg number EIT.

\begin{figure}[!ht]
\centering
\includegraphics[scale=0.85,trim=30 530 30 52,clip,tics=10]{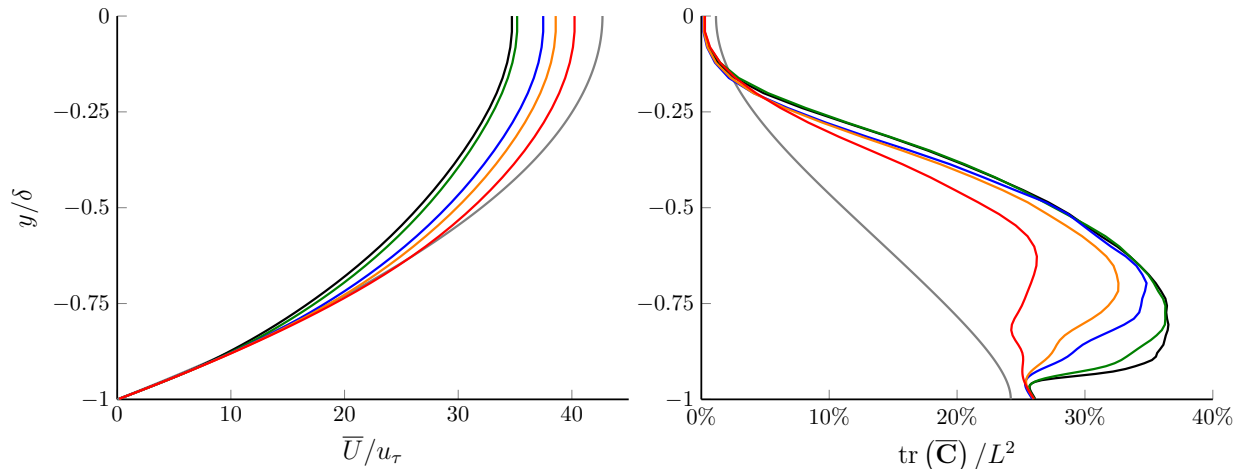}
\caption{Mean streamwise velocity profile (left) and mean polymer elongation profile (right) for a two-dimensional flow at $\mathrm{Wi}_\tau$ = 40 and different Schmidt numbers: \textcolor{gray}{$\mathrm{Sc}$ = 0.5 (laminar)}, \textcolor{red}{$\mathrm{Sc}$ = 9}, \textcolor{orange}{$\mathrm{Sc}$ = 16}, \textcolor{blue}{$\mathrm{Sc}$ = 25}, \textcolor{black!50!green}{$\mathrm{Sc}$ = 100} and $\mathrm{Sc} = \infty$.}
\label{fig:Statistics}
\end{figure}

The effects of the Schmidt number on the different scales of the turbulence can be evaluated in the right sub-figure of Figure~\ref{fig:Spectra}. For the present conditions, a Schmidt number $\mathrm{Sc} <$ 9 leads to a complete laminarization of the flow. For $\mathrm{Sc} \ge$ 9, the range of scales affected is directly related to the amount of polymer diffusion introduced. A Schmidt number of $\mathrm{Sc}$ = 100 does not have a significant impact on the large scales and only damps a part of the sub-Kolmogorov scales. However, a wider range of scales is affected by the diffusion as the Schmidt number decreases. For values 9 $\le \mathrm{Sc} <$ 100, all the scales are diffused and the turbulent kinetic energy spectrum progressively narrows and weakens. On top of that, the quantitative effects of unrealistically large polymer diffusion on the mean velocity and polymer elongation profiles can be appreciated in Figure~\ref{fig:Statistics}.
Lowering the Schmidt number increases the mean flow velocity and reduces the mean polymer extension. The diffusion tends to homogenize the elastic stress spatial distribution and counterbalance the stretching mechanism. Thus, diffusing the small-scale polymeric stress fluctuations of the sheets hinders the turbulence generation mechanism leading to a reduction of the velocity fluctuations magnitude. As a consequence, the turbulent flow resistance is weakened and the flow-rate increases. It is worth noting that diffusion has a double effect on the polymer elongation. In addition to directly diffusing the highly-stretched sheets, the reduced level of turbulence in the flow induces a depletion of the stretching mechanism by the solvent inertia. 

More generally, the excessive polymer diffusion tends to prevent elastic fluctuations to develop and therefore, inhibits the elasto-inertial instability. As long as the diffusion is not too strong, the polymer stress gradients remain large enough to sustain the solvent chaotic motion necessary to regenerate the small-scale elastic sheets and preserve the global EIT dynamics.

\section{\label{sec:Conclusion}Conclusion}
The key outcomes of this investigation are that (i) EIT is fundamentally two-dimensional, (ii) its inherent existence in three-dimensional flows overwhelms the dynamics providing that the polymer solution is able to inhibit the growth of near-wall vortices through the polymer drag reduction mechanism, and (iii) the mechanism of EIT is small scales as shown by the Schmidt number study. To further demonstrate outcome (ii), the present results can be compared to those of \citet{Wang2014}, who used the same Reynolds number, other polymer parameters and a Schmidt number $\mathrm{Sc}=0.5$. They reported intermediate drag reduction at $\mathrm{Wi}_\tau=40$ and MDR at $\mathrm{Wi}_\tau=100$, yet with flow structures that are clearly of Newtonian nature. Our Schmidt number study demonstrates that EIT ceases to exist when $\mathrm{Sc}<9$, yet for $\mathrm{Sc}=\infty$, streamwise vortices (a distinct feature of Newtonian turbulence) are only observed intermittently and only for $\mathrm{Wi}_\tau=40$. For $\mathrm{Wi}_\tau=100$, EIT is strong enough to fully dominate the dynamics of the flow. 

The introduction of large polymer diffusion into the system damps a significant part of the elastic scales that are necessary to feed turbulence, eventually leading to the flow laminarization. The transition to the laminar state does not systematically occur as EIT can withstand a low/moderate amount of diffusion without inhibiting the turbulence generation mechanism. Nevertheless, adding unrealistically large polymer diffusion to the system has an important impact on the dynamics and leads to significant quantitative changes, even for Schmidt numbers as large as 10$^2$. This observation demonstrates the importance of the small elastic scales and highlights the necessity of preserving the hyperbolic nature of the transport equation by using zero-diffusion polymers. Outcome (iii) is critical for the understanding of polymer drag reduction and specifically MDR. Previous studies have concluded that the nature of MDR is Newtonian \cite{Xi2010,Xi2010a,Xi2012,Wang2014,wang2017} for the same marginally critical Reynolds number. However, the present results suggest that their conclusion is biased by the use of numerical methods that dissipate the small-scale dynamics critical to EIT. Our study concludes that, for the Reynolds number considered here, MDR is EIT.

The extension to higher Reynolds number is not straightfoward and will be the focus of future investigations. The authors speculate that EIT should dominate MDR, providing that polymer parameters are adequate in the range of Reynolds numbers for which the sustaining mechanism of wall turbulence is confined to the inner region, as shown by \citet{Jimenez1999}. For high Reynolds numbers, energy transfers between outer and inner regions could compete and possibly dominate EIT.

\section*{\label{sec:Acknowledgments}Acknowledgments}
S. Sid is grateful to the Fonds de la Recherche Scientifique (F.R.S.-FNRS) for funding the present research project. V.E. Terrapon acknowledges the financial support of a Marie Curie FP7 Career Integration Grant [grant number PCIG10-GA-2011-304073] and of a grant from the 'Fonds sp\'{e}ciaux pour la recherche' of the University of Liege (Cr\'{e}dits classiques 2013) [grant number C-13/19]. Computational resources have been provided by the Consortium des \'{E}quipements de Calcul Intensif (CECI) funded by the Fonds de la Recherche Scientifique (F.R.S.-FNRS), the Vermont Advanced Computing Center (VACC) and the Partnership for Advanced Computing in Europe (PRACE) that are appreciatively acknowledged. The present research also benefited from computational resources made available on the Tier-1 supercomputer of the F\'{e}d\'{e}ration Wallonie-Bruxelles, infrastructure funded by the Walloon Region under the grant agreement n$^{\text{o}}$117545. Y. Dubief's contribution is supported by a National Science Foundation/Department of Energy Partnership Award No: 1258697. 

\bibliography{bibli}

\newpage
\section*{\label{sec:SupplMat}Supplementary Material}
The goal of this supplementary material is to demonstrate that the simulation methodology used to study the physics of EIT is appropriate and provides accurate results. To do so, two codes using different numerics have been developed and compared. In addition, a mesh convergence analysis has been carried out to determine the size of the smallest physical scale that should be captured to properly simulate viscoelastic turbulence. This convergence analysis is also used to measure the effects of the spatial resolution on the solution.

In order to emphasize that our conclusions are also valid for more challenging cases, the flow conditions considered in this analysis slightly differ from those reported in the main paper. Specifically, a lower Reynolds number, $\mathrm{Re}_\tau$ = 40, and a larger Weissenberg number, $\mathrm{Wi}_\tau$ = 310, have been selected. This reduces the amount of inertia in the flow and thus amplifies the effects of polymer diffusion on the solution, while keeping a self-sustained EIT state. All the results reported in this supplementary material are obtained from two-dimensional simulations.

\begin{figure}[!ht]
\centering
\includegraphics[scale=1,trim=80 550 80 50,clip,tics=10]{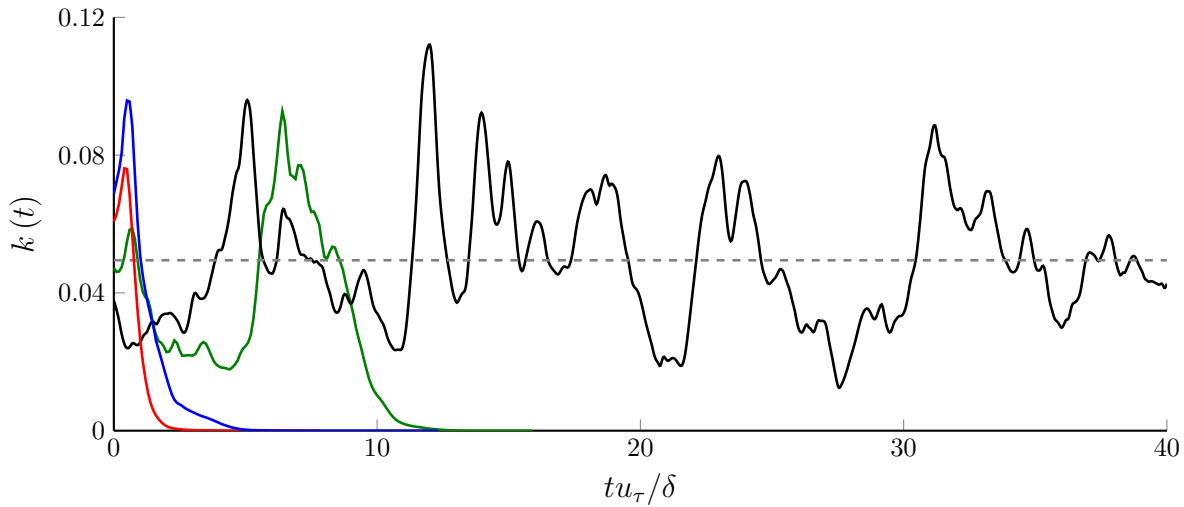}
\caption{Time series of the volume-averaged instantaneous turbulent kinetic energy $k\left( t \right)$ for $\mathrm{Re}_\tau$ = 40, $\mathrm{Wi}_\tau$ = 310 and different Schmidt numbers: \textcolor{red}{$\mathrm{Sc} = 9$}, \textcolor{blue}{$\mathrm{Sc} = 25$}, \textcolor{black!50!green}{$\mathrm{Sc} = 50$}, $\mathrm{Sc} = 100$. The initial time, $t=0$, is defined as the time when the blowing-suction perturbation is turned off.}
\label{fig:fullyResolved_TimeSeries}
\end{figure}
Time series of the volume-averaged instantaneous turbulent kinetic energy $k\left( t \right)$ are reported in Figure~\ref{fig:fullyResolved_TimeSeries} for different Schmidt numbers. The quantity $k\left( t \right)$ is defined as
\begin{equation}
	k \left( t \right)  = \left\langle \frac{1}{2} \frac{{\mathbf{u}^\prime}\left(\mathbf{x},t\right)  \cdot \mathbf{u}^\prime \left(\mathbf{x},t\right)}{u_\tau^2}\right\rangle_\Omega \qquad \text{with} \qquad \mathbf{u}^\prime \left(\mathbf{x},t\right) = \mathbf{u} \left(\mathbf{x},t\right) - \overline{\mathbf{u} \left(\mathbf{x},t\right)} \,,
\end{equation}
where the spatial and time averaging operators,
\begin{equation}
	\left\langle \bullet \right\rangle_\Omega = \frac{1}{\Omega} \int_\Omega \bullet \ \mathrm{d}\Omega \qquad \text{and} \qquad \bar{\bullet} = \frac{1}{T_s} \int_{t_0}^{t_0+T_s} \bullet \ \mathrm{d}t \,,
\end{equation}
are introduced, $\Omega$ is the channel volume and $T_s$ the sampling window for temporal averaging. The evolution of the turbulent kinetic energy shows that, for the present flow conditions, $\mathrm{Sc}$ has to be between 50 and 100 to allow EIT to sustain itself over time; large polymer diffusivity leads to flow laminarization. In this case, the critical Schmidt number above which a non-laminar state is sustained is larger than the one reported for $\mathrm{Re}_\tau$ = 85, $\mathrm{Wi}_\tau$ = 40 owing to the lower contribution of inertia at $\mathrm{Re}_\tau$ = 40. 

\begin{figure}[!ht]
\centering
\includegraphics[scale=1,trim=50 510 50 52,clip,tics=10]{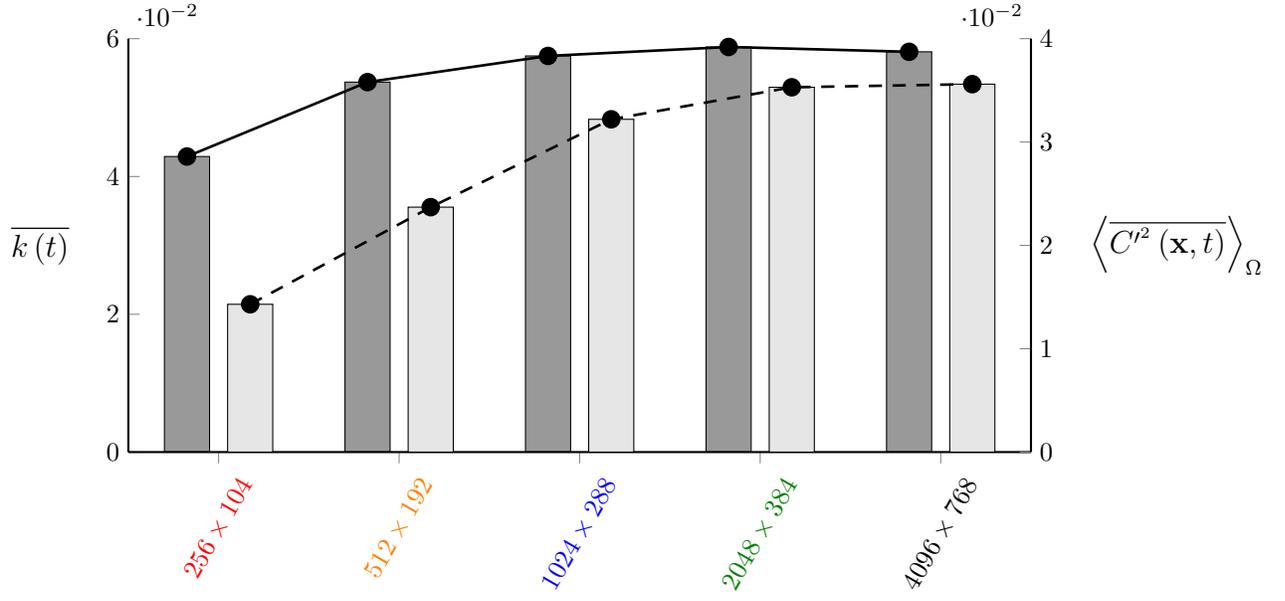}
\caption{Mean volume-averaged turbulent kinetic energy $\overline{k\left(t\right)}$ (dark gray bars, plain line, left $y$-axis) and mean volume-averaged polymer elongation mean square fluctuations $\left\langle\overline{C^{{\prime}^2}\left(\mathbf{x},t\right)} \right\rangle_\Omega$ (light gray bars, dashed line, right $y$-axis) for $\mathrm{Re}_\tau$ = 40, $\mathrm{Wi}_\tau$ = 310, $\mathrm{Sc}$ = 100 using different spatial resolutions.}
\label{fig:Convergence_k}
\end{figure}

Each simulation reported in Figure~\ref{fig:fullyResolved_TimeSeries} has been performed using a spatial resolution $\Delta$ that captures the viscous Batchelor scale, i.e., $\Delta \approx \eta_{\text{B},\nu}= \delta_\nu \mathrm{Sc}^{-1/2} = \delta \ \mathrm{Re}_\tau^{-1} \mathrm{Sc}^{-1/2}$. Because (i) Batchelor theory only applies to a passive scalar, which is not the case in EIT, and (ii) EIT strongly differs from classical Newtonian turbulence, the pertinence of defining the mesh based on $\delta_\nu$ and $\eta_\text{B}$ is not \emph{a priori} obvious. In order to verify \emph{a posteriori} how the smallest physical scale compares to $\eta_{\text{B},\nu}$, a mesh convergence analysis has been performed. The mean volume-averaged turbulent kinetic energy $\overline{k\left( t \right)}$ and mean volume-averaged polymer elongation mean square fluctuations $\left\langle\overline{C^{{\prime}^2}\left(\mathbf{x},t\right)} \right\rangle_\Omega$ are reported in Figure~\ref{fig:Convergence_k} for different meshes, where $C^\prime \left(\mathbf{x},t\right) = \mathrm{tr} \left( \mathbf{C} \left(\mathbf{x},t\right) - \overline{\mathbf{C} \left(\mathbf{x},t\right)} \right)/L^2$. 
The solution obtained for $\Delta \approx \eta_{\text{B},\nu}$ (i.e., on the 4096$\times$768 mesh) is well-converged, and even a 1024$\times$288 mesh provides a relatively accurate solution. For coarser resolutions, the level of turbulent fluctuations starts to rapidly decrease, indicating that the mesh is not able to capture all the relevant scales contributing to the flow dynamics. 

\begin{figure}[!ht]
\centering
\includegraphics[scale=1,trim=55 540 50 52,clip,tics=10]{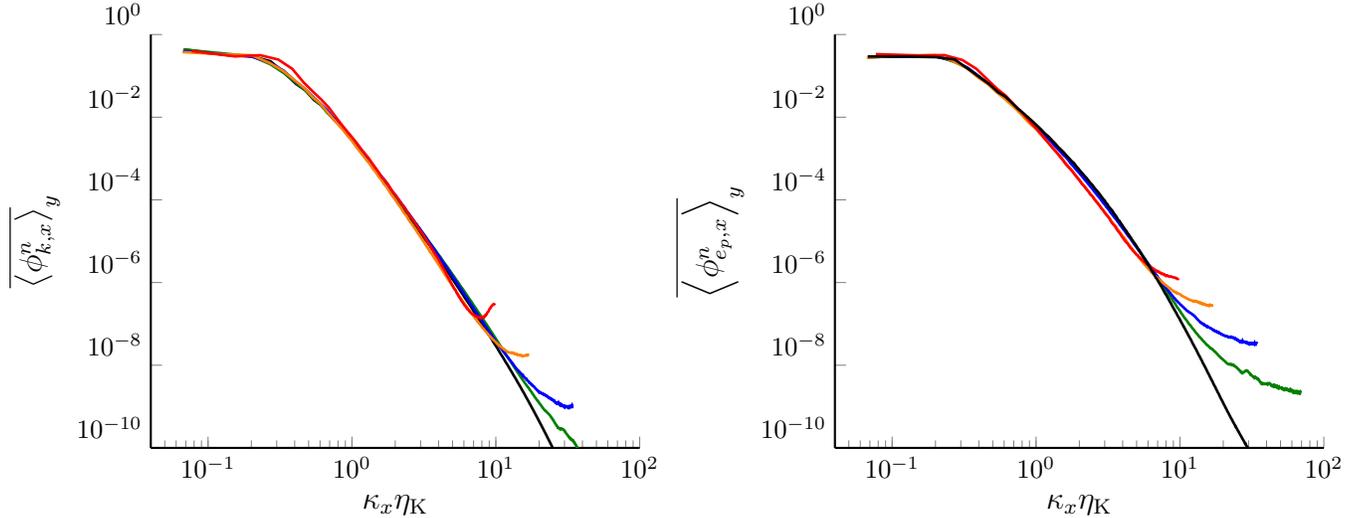}
\caption{Wall-normal and time-averaged normalized streamwise spectra of the turbulent kinetic energy $k$ and elastic energy $e_p$ for $\mathrm{Re}_\tau$ = 40, $\mathrm{Wi}_\tau$ = 310, $\mathrm{Sc}$ = 100 on meshes with \textcolor{red}{256$\times$104}, \textcolor{orange}{512$\times$192}, \textcolor{blue}{1024$\times$288}, \textcolor{black!50!green}{2048$\times$384} and 4096$\times$768 grid points. Wavenumbers are pre-multiplied by the computed wall Kolmogorov length scale $\eta_\text{K} = \left( \nu^3 / \epsilon_\text{wall} \right)^{1/4}$.}
\label{fig:Convergence_spectra}
\end{figure}
The distribution of fluctuations across spatial scales is analyzed though the wall-normal and time-averaged normalized streamwise spectra of the turbulent kinetic energy $k$ and elastic energy $e_p$, as shown in Figure~\ref{fig:Convergence_spectra} for different mesh sizes. The two normalized spectra are respectively defined as
\begin{eqnarray}
	\phi_{k,x}^n \left( \kappa_x, y, t \right) = \frac{1}{I_{\phi_k}} \phi_{k,x} \left( \kappa_x, y, t \right) \qquad &\text{with}& \qquad \phi_{k,x} \left( \kappa_x, y, t \right) = \frac{1}{2 u_\tau^2} \left[ \hat{\mathbf{u}} \left( \kappa_x, y, t \right) \cdot \hat{\mathbf{u}}^* \left( \kappa_x, y, t  \right) \right] \,, \\
	\phi_{e_p,x}^n \left( \kappa_x, y, t \right) = \frac{1}{I_{\phi_{e_p}}} \phi_{e_p,x} \left( \kappa_x, y, t \right) \qquad &\text{with}& \qquad \phi_{e_p,x} \left( \kappa_x, y, t \right) = \hat{C^\prime} \left( \kappa_x, y, t \right) \cdot \hat{C^\prime}^* \left( \kappa_x, y, t  \right) \,,
\end{eqnarray}
where $I_{\phi_\bullet}$ is the integral of $\overline{\left\langle \phi_{\bullet,x} \right\rangle}_y$ over $-\infty < \kappa_x < \infty$, the hat operator stands for the Fourier transform along the streamwise $x$-direction, $\hat{\bullet} = \mathcal{F}_x \left( \bullet \right)$, and the star superscript denotes the complex conjugate. Although the integrals $I_{\phi_k}$ and $I_{\phi_{e_p}}$ strongly depend on the discretization as discussed previously, the normalized spectral densities remain remarkably similar on the different meshes. This suggests that the physical mechanism responsible for the creation of turbulence remains mostly unaffected by the mesh discretization. The effect of a poor spatial resolution manifests itself mostly by the global reduction of the turbulent intensity across all scales. It is worth noting that the 1024$\times$288 mesh, i.e., the coarsest discretization still providing a relatively accurate solution, is also the coarsest mesh that properly captures a scale that is $\sqrt{\mathrm{Sc}}$ = 10 times smaller than the computed wall Kolmogorov length scale $\eta_\text{K} = \left( \nu^3 / \epsilon_\text{wall} \right)^{1/4}$. This suggests that Batchelor theory for passive scalars provides a good estimate of the smallest scales, even for an active scalar like in EIT. Note finally that the computed wall Kolmogorov scale $\eta_\text{K}$ is about three times larger than the viscous length scale $\delta_\nu$. 

\begin{figure}[!ht]
\centering
\includegraphics[scale=1,trim=55 540 0 52,clip,tics=10]{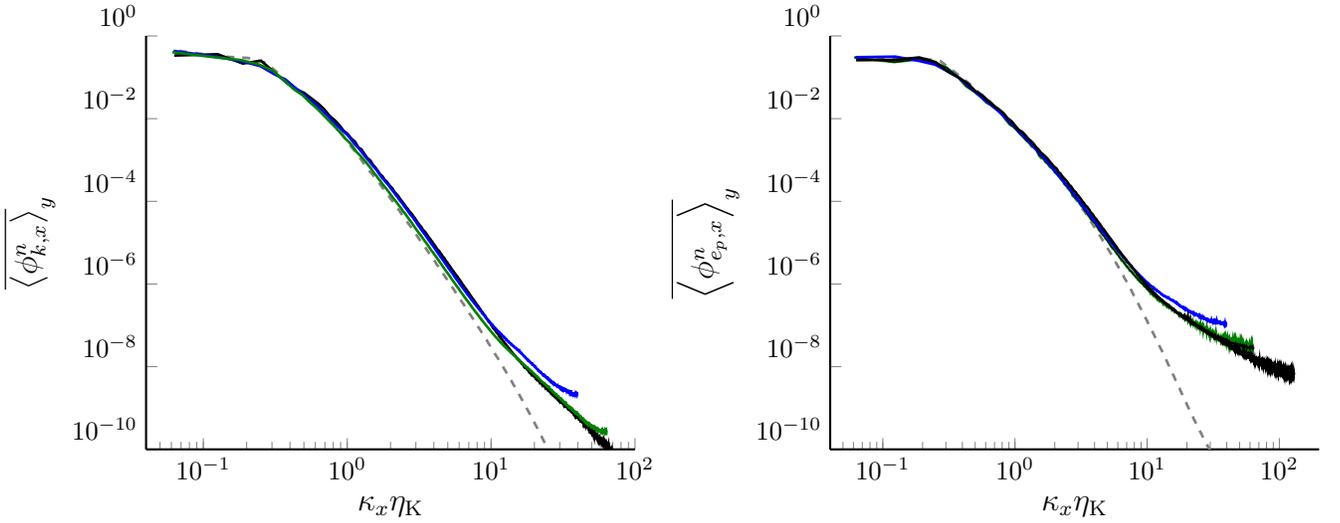}
\caption{Wall-normal and time-averaged normalized streamwise spectra of the turbulent kinetic energy $k$ and elastic energy spectra $e_p$ for different discretizations and Schmidt numbers at $\mathrm{Re}_\tau$ = 40, $\mathrm{Wi}_\tau$ = 310: \textcolor{gray}{$\mathrm{Sc}$ = $100$, 4096$\times$768 (dashed)}; \textcolor{blue}{$\mathrm{Sc}$ = $\infty$, 1280$\times$384}; \textcolor{black!50!green}{$\mathrm{Sc}$ = $\infty$, 2080$\times$480} and $\mathrm{Sc}$ = $\infty$, 4096$\times$768. Wavenumbers are pre-multiplied by the computed wall Kolmogorov length scale $\eta_\text{K} = \left( \nu^3 / \epsilon_\text{wall} \right)^{1/4}$.}
\label{fig:Hyperbolic_Spectra}
\end{figure}

Finally, the diffusive effect of a finite Schmidt number equal to 100 is evaluated by comparison with simulations using $\mathrm{Sc} = \infty$. As zero-diffusion polymers should theoretically lead to infinitely small scales, the effect of the finite grid used in simulations is assessed by considering three different grid resolutions. The normalized turbulent kinetic and elastic energy spectra are shown in Figure~\ref{fig:Hyperbolic_Spectra}.
It can be observed that the smallest scales for $\mathrm{Sc}$ = $\infty$ on finite meshes are slightly affected by numerical errors, as the diffusive cut-off is not physical but numerical, dictated by the finite size of the smallest grid cells. Part of the energy of the unresolved scales thus piles-up at the end of the spectrum. However, these numerical errors are small and mainly confined at the small scales. Moreover, refining the mesh reduces the impact of these numerical errors. Note that the spectra are here normalized by their respective integrated value. The integrals $I_{\phi_k}$ and $I_{\phi_{e_p}}$ are therefore reported in table~\ref{table:energyIntegrals} for the three cases.
\begin{table}[h!]
\caption{Turbulent kinetic energy integral $I_{\phi_k}$ and fluctuating elastic energy integral $I_{\phi_{e_p}}$ for $\mathrm{Re}_\tau$ = 40, $\mathrm{Wi}_\tau$ = 310.}
\begin{ruledtabular}
\begin{tabular}{lcccc}
  & $n_x \times n_y$ & $\mathrm{Sc}$ & $I_{\phi_k}$ & $I_{\phi_{e_p}}$\\[.3em]
  \hline \\[-.6em]
  \textcolor{gray}{A} & 4096$\times$768 & 100 & 1.53 $\cdot$ 10$^{-2}$ & 1.05 $\cdot$ 10$^{-2}$ \\
  \hline \\[-.6em]
  \textcolor{blue}{B} & 1280$\times$384 & $\infty$ & 1.53 $\cdot$ 10$^{-2}$ & 0.87 $\cdot$ 10$^{-2}$ \\
  \textcolor{black!50!green}{C} & 2080$\times$480 & $\infty$ & 1.62 $\cdot$ 10$^{-2}$ & 0.88 $\cdot$ 10$^{-2}$ \\
  D & 4096$\times$768 & $\infty$ & 1.68 $\cdot$ 10$^{-2}$ & 1.04 $\cdot$ 10$^{-2}$
\end{tabular}
\end{ruledtabular}
\label{table:energyIntegrals}
\end{table}
The amount of turbulent kinetic energy in the flow is larger for $\mathrm{Sc}$ = $\infty$ compared to a finite Schmidt number of 100, indicating that a Schmidt number as large as $10^2$ has still a measurable diffusive effect on the solution for the flow conditions considered. Additionally, a finer spatial resolution significantly increases the fluctuations of both the velocity and polymer extension, as a larger portion of the spectrum is captured. This again highlights the importance of small scales in the dynamics of EIT. 

\begin{figure}[!ht]
\centering
\includegraphics[scale=1,trim=50 540 50 52,clip,tics=10]{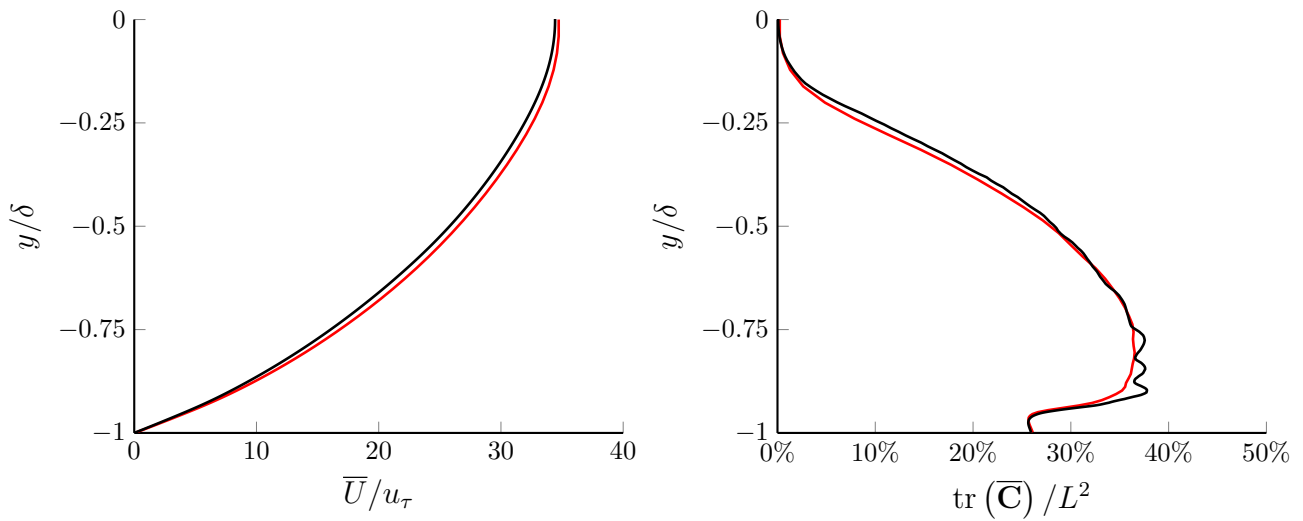}
\caption{Mean profiles of the streamwise velocity (left) and polymer elongation (right) for $\mathrm{Re}_\tau$ = 85, $\mathrm{Wi}_\tau$ = 40, $\mathrm{Sc} = \infty$. The black profiles are obtained with code A using a mesh with 1024$\times$288 grid points whereas the red profiles are obtained with code B using a mesh with 512$\times$128 grid points.}
\label{fig:Code_Comparison}
\end{figure}

The behavior of the solution for $\mathrm{Sc} =\infty$ on  grids of finite size results from the dissipation and dispersion introduced by the numerical scheme, similarly to what is observed around shocks in supersonic flows. The adequate grid size is thus challenging to estimate \emph{a priori}. This is further illustrated by comparing global flow statistics obtained with two codes using different numerical schemes. The mean profiles of streamwise velocity and polymer elongation are shown in Figure~\ref{fig:Code_Comparison} for $\mathrm{Re}_\tau$ = 85, $\mathrm{Wi}_\tau$ = 40, $\mathrm{Sc} = \infty$.
The two codes mainly differ in their discretization of the advective term $\mathbf{u} \cdot \nabla \mathbf{C}$ in the conformation tensor transport equation. This term is critical, as it is responsible for the creation of small elastic scales that are necessary to sustain elasto-inertial turbulence. Code A features a third-order WENO scalar interpolation \citep{Liu2013} on a staggered grid, while code B relies on a fourth-order compact interpolation scheme on a collocated grid. Because the third-order WENO scheme is more dissipative, code A requires a finer mesh than code B for converged results, as shown in Figure~\ref{fig:Code_Comparison}. Nonetheless, if an adequate grid resolution is used, both codes lead to the same mean profiles. 

Overall, this supplementary material demonstrates that:
\begin{itemize}
	\item the critical Schmidt number below which the flow becomes laminar depends on the flow conditions; in particular, elasto-inertial turbulence at larger Reynolds number remains turbulent for lower Schmidt numbers;
	\item the Batchelor length scale $\eta_\text{B} = \eta_\text{K} \mathrm{Sc}^{-1/2}$ is a good estimation of the smallest elastic scale for finite Schmidt numbers;
	\item a poor spatial resolution reduces the turbulence intensity in the flow but does not significantly modify its dynamical behaviour;
	\item the flow solution obtained using zero-diffusion polymers ($\mathrm{Sc} = \infty$) converges as the spatial resolution is increased; 
	\item the discretization necessary to obtain an accurate solution for $\mathrm{Sc} = \infty$ depends on the flow conditions and on the numerical schemes.
\end{itemize}
These conclusions suggest that the important observations described in the main article are not artefacts of the numerical approach, but indeed represent the actual physics of such flows. 
\end{document}